\ifx\stupidformat\undefined
\documentclass[12pt,a4paper,twoside]{artikel1}
\else
\documentclass[12pt,a4paper]{artikel1}
\newcommand\auname{\markright{G. J. van Oldenborgh, G. Burgers and A. Klein Tank}}
\usepackage[noheads,tablesfirst]{endfloat}
\usepackage{doublespace}
\AtBeginDelayedFloats{}
\auname
\fi
\usepackage{beton}
\usepackage{euler}
\usepackage{epsfig}
\usepackage[round]{natbib}
\newcommand\dg{${}^\circ$}
\newcommand{\nop}[1]{}

\newlength\figwidth
\begin{document}
\title{On the El-Ni\~no Teleconnection to Spring Precipitation in Europe}
\author{%
Geert Jan van Oldenborgh\\
Gerrit Burgers\\
Albert Klein Tank\\[1ex]
\textit{KNMI, De Bilt, Netherlands}}
\date{December 1998}
\maketitle
\ifx\stupidformat\undefined\else\clearpage\fi
\begin{abstract}
In a statistical analysis of more than a century of data we find a
strong connection between strong warm El Ni\~no winter events and high
spring precipitation in a band from Southern England eastwards into
Asia.  This relationship is an extension of the connection mentioned
by \citet{KiladisDiaz1989}, and much stronger than the winter season
teleconnection that has been the subject of other studies.  Linear
correlation coefficients between DJF NINO3 indices and MAM
precipitation are higher than $r=0.3$ for individual stations, and as
high as $r=0.49$ for an index of precipitation anomalies around 50\dg
N from 5\dg W to 35\dg E{}.  The lagged correlation suggests that
south-east Asian surface temperature anomalies may act as intermediate
variables.
\end{abstract}
\ifx\stupidformat\undefined\else
\vspace{\fill}
\clearpage\fi


\section{Introduction}

The recent strong El Ni\~no has stirred up again interest in possible
teleconnections to Europe.  A few influences have been mentioned in
previous studies.  During the winter season, Scandinavia was found to
be colder and dryer just after cold (La Ni\~na) events
\citep{Berlage1966}, and an increase in temperature and precipitation
with El Ni\~no conditions was found in this area
\citep{VanLoonMadden1981,HalpertRopelewski1992}.  Central
European winters would have the opposite tendencies: warmer and wetter
\citep{KiladisDiaz1989} during ENSO warm events; this squares with the
increase of cyclonic Gro{\ss}wetter days observed in Germany
\citep{Fraedrich1994} and England \citep{Wilby1993} during El Ni\~no
years.

For the spring season after an El Ni\~no event \citep{KiladisDiaz1989}
note precipitation anomalies in central Europe, and
\citep{HalpertRopelewski1992} find temperature anomalies in
South-western Europe and Northern Africa.

Recently, new data sets of historical data have become available.
This has opened up the possibility to check existing conjectures at
higher statistical significance levels.  This study was initiated by a
search for an ENSO influence in the Netherlands through tropical storm
activity.  Tropical storm and hurricane activity on the Atlantic is
suppressed by El Ni\~no \citep{Gray1984a}, and many catastrophic
downpours in De Bilt are remnants of tropical storms during the
Atlantic hurricane season.  Unfortunately, this does not lead to an
observable anti-correlation between the NINO3 index and precipitation
(or high-precipitation events).

In this article we present a new analysis of the strongest statistical
connection between ENSO and the weather in Europe we did find:
increased spring precipitation after an El Ni\~no event.  In
section~\ref{sec:debilt} we detail the connection for one station, De
Bilt in the Netherlands.  The relationship is extended over Europe in
section~\ref{sec:europe}, and we search for possible mechanisms in
section~\ref{sec:mechanisms}.


\section{Rain in De Bilt}
\label{sec:debilt}

Two years ago we noticed that there was a correlation between the
strength of an El Ni\~no, quantified by the NINO3.4
index\ifx\stupidformat\undefined\footnote{the
NINO3.4 index is\else\ (\fi the average sea surface temperature in the region
5\dg S--5\dg N, 120\dg W--170\dg W, the values where obtained from
NOAA/NCEP
\citep{ReynoldsAnalyses}\ifx\stupidformat\undefined.}\else)\fi\ 
and spring (MAM) precipitation in
De Bilt (central Netherlands) using data from 1950 to 1995.  With a
3-month lag the linear correlation coefficient was 0.30, with a
nominal significance of 95\%.  Given that we had considered more than
24 possible relationships, this result was not very convincing.  It
hinged on one extreme event: in 1983 the spring had been
extraordinarily wet (see Fig.~\ref{fig:scatter}).  Recently we used
the precipitation series back to 1849 and forward to
1998\ifx\stupidformat\undefined\footnote{\else\ (\fi the
series is available upon request from the
KNMI\ifx\stupidformat\undefined.}\else)\fi\ including a few more
strong El Ni\~no events.  Also, \citet{Kaplan1998} give a
reconstruction of the NINO3
index\ifx\stupidformat\undefined\footnote{the NINO3 index 
is \else\ (\fi the average sea surface temperature in the region 
5\dg S--5\dg N, 90\dg W--150\dg
W\ifx\stupidformat\undefined.}\else)\fi\ from 
1856 to 1991.  It uses only sea surface temperature 
(SST) measurements, and correlates quite well with the Jakarta SO
index of \citet{Konnen1998}: $r=0.66$ in DJF and $0.67$ in MAM over
133 years in 1859--1996.  From 1950 onwards we use the NCEP analyses
\citep{ReynoldsAnalyses}, over the overlap period the correlation with
the Kaplan NINO3 is $0.97$.  An analysis using the standard SO index
\citep{Allan1991,Konnen1998} gives essentially the same results.

Figure~\ref{fig:scatter} shows that the relationship between the
winter (DJF) NINO3 index and spring (MAM) precipitation in De Bilt was
confirmed in the new analysis.  The linear correlation coefficient is
$r=0.35$, nominally this has a chance $P<10^{-4}$ of being random.
The average of the four Dutch stations with data from 1867 (De Bilt, 
Groningen, Den Helder and Hoofddorp) gives a 
correlation of $0.40$.

\begin{figure*}
\begin{center}
\unitlength 1.5cm
\begin{picture}(6.4,7.4)(-2.4,-0.4)
\thicklines
\put(-2,0){\line(1,0){6}}
\put(-2,0){\line(0,1){7}}
\put(-2,7){\line(1,0){6}}
\put(4,0){\line(0,1){7}}
\put(3.5,0.4){\makebox(0,0)[tr]{$r=0.35$}}
\thinlines 
\put(-2,1.682){\line(1,0){6}}
\put(-2,2.569){\line(1,0){6}}
\put(-2,3.374){\line(1,0){6}}
\put(-2,4.147){\line(1,0){6}}
\put(-1,0){\line(0,1){.10}}
\put( 0,0){\line(0,1){.10}}
\put( 1,0){\line(0,1){.10}}
\put( 2,0){\line(0,1){.10}}
\put( 3,0){\line(0,1){.10}}
\put(-1,7){\line(0,-1){.10}}
\put( 0,7){\line(0,-1){.10}}
\put( 1,7){\line(0,-1){.10}}
\put( 2,7){\line(0,-1){.10}}
\put( 3,7){\line(0,-1){.10}}
\put(-2,1){\line(1,0){.10}}
\put(-2,2){\line(1,0){.10}}
\put(-2,3){\line(1,0){.10}}
\put(-2,4){\line(1,0){.10}}
\put(-2,5){\line(1,0){.10}}
\put(-2,6){\line(1,0){.10}}
\put(-2,7){\line(1,0){.10}}
\put(4,1){\line(-1,0){.10}}
\put(4,2){\line(-1,0){.10}}
\put(4,3){\line(-1,0){.10}}
\put(4,4){\line(-1,0){.10}}
\put(4,5){\line(-1,0){.10}}
\put(4,6){\line(-1,0){.10}}
\put(-2,-.05){\makebox(0,0)[t]{$-2$}}
\put(-1,-.05){\makebox(0,0)[t]{$-1$}}
\put( 0,-.05){\makebox(0,0)[t]{$0$}}
\put( 1,-.05){\makebox(0,0)[t]{$+1$}}
\put( 2,-.05){\makebox(0,0)[t]{$+2$}}
\put( 3,-.05){\makebox(0,0)[t]{$+3$}}
\put( 4,-.05){\makebox(0,0)[t]{$+4$}}
\put( 4,-.35){\makebox(0,0)[tr]{DJF NINO3}}
\put(-2.05,0){\makebox(0,0)[r]{$0$}}
\put(-2.05,2){\makebox(0,0)[r]{$100$}}
\put(-2.05,4){\makebox(0,0)[r]{$200$}}
\put(-2.05,6){\makebox(0,0)[r]{$300$}}
\put(-2.05,7){\makebox(0,0)[rt]{\shortstack{MAM prec.\\{}[mm]}}}
\put( -.5733,1.980){\makebox(0,0){$\underline{57}$}}
\put( -.0600,1.528){\makebox(0,0){$\underline{58}$}}
\put( -.3500,3.806){\makebox(0,0){$\underline{59}$}}
\put( -.6400,3.938){\makebox(0,0){$\underline{60}$}}
\put( -.5533,3.244){\makebox(0,0){$\underline{61}$}}
\put( -.3767,1.562){\makebox(0,0){$\underline{62}$}}
\put( -.4933,1.662){\makebox(0,0){$\underline{63}$}}
\put(  .0267,1.740){\makebox(0,0){$\underline{64}$}}
\put(  .8533,1.940){\makebox(0,0){$\underline{65}$}}
\put( 1.2300,2.582){\makebox(0,0){$\underline{66}$}}
\put(  .7433,2.140){\makebox(0,0){$\underline{67}$}}
\put(  .4600,2.744){\makebox(0,0){$\underline{68}$}}
\put( 1.1133,3.856){\makebox(0,0){$\underline{69}$}}
\put( -.8400,1.992){\makebox(0,0){$\underline{70}$}}
\put( -.4033,2.056){\makebox(0,0){$\underline{71}$}}
\put( -.4200,2.370){\makebox(0,0){$\underline{72}$}}
\put( -.7867,2.576){\makebox(0,0){$\underline{73}$}}
\put(-1.0767,3.272){\makebox(0,0){$\underline{74}$}}
\put( -.6867,1.698){\makebox(0,0){$\underline{75}$}}
\put( -.8567,3.720){\makebox(0,0){$\underline{76}$}}
\put(  .5767,2.800){\makebox(0,0){$\underline{77}$}}
\put( 2.6567,4.396){\makebox(0,0){$\underline{78}$}}
\put( -.1100,2.568){\makebox(0,0){$\underline{79}$}}
\put( -.8267,1.564){\makebox(0,0){$\underline{80}$}}
\put(  .3867,3.648){\makebox(0,0){$\underline{81}$}}
\put( -.4033,3.720){\makebox(0,0){$\underline{82}$}}
\put( -.5633,1.632){\makebox(0,0){$\underline{83}$}}
\put( -.2300,1.654){\makebox(0,0){$\underline{84}$}}
\put(  .6300,2.452){\makebox(0,0){$\underline{85}$}}
\put(  .1000,2.960){\makebox(0,0){$\underline{86}$}}
\put( -.9733,2.492){\makebox(0,0){$\underline{87}$}}
\put(  .4900,2.994){\makebox(0,0){$\underline{88}$}}
\put( 1.9400,3.302){\makebox(0,0){$\underline{89}$}}
\put(-1.1133,2.934){\makebox(0,0){$\underline{90}$}}
\put( -.0033,3.240){\makebox(0,0){$\underline{91}$}}
\put( -.0100,1.432){\makebox(0,0){$\underline{92}$}}
\put(-1.3933, .928){\makebox(0,0){$\underline{93}$}}
\put( -.6767,2.870){\makebox(0,0){$\underline{94}$}}
\put( -.3567,3.156){\makebox(0,0){$\underline{95}$}}
\put(  .5167,1.936){\makebox(0,0){$\underline{96}$}}
\put( 1.8133,3.788){\makebox(0,0){$\underline{97}$}}
\put( -.3733,3.518){\makebox(0,0){$\underline{98}$}}
\put( -.5567,3.950){\makebox(0,0){$\underline{99}$}}
\put( 1.4400,2.318){\makebox(0,0){$\underline{00}$}}
\put(  .3267,3.766){\makebox(0,0){$01$}}
\put(  .2233,3.166){\makebox(0,0){$02$}}
\put( 1.1767,5.012){\makebox(0,0){$03$}}
\put( -.6833,2.482){\makebox(0,0){$04$}}
\put(  .9467,3.304){\makebox(0,0){$05$}}
\put( 1.2667,3.352){\makebox(0,0){$06$}}
\put( -.4033,3.328){\makebox(0,0){$07$}}
\put( -.0067,2.588){\makebox(0,0){$08$}}
\put( -.8600,3.836){\makebox(0,0){$09$}}
\put( -.9733,2.970){\makebox(0,0){$10$}}
\put( -.6200,2.094){\makebox(0,0){$11$}}
\put( 1.3600,3.576){\makebox(0,0){$12$}}
\put(  .2300,3.496){\makebox(0,0){$13$}}
\put(  .9433,4.456){\makebox(0,0){$14$}}
\put( 1.4900,3.572){\makebox(0,0){$15$}}
\put( -.4833,4.642){\makebox(0,0){$16$}}
\put(-1.2933,1.926){\makebox(0,0){$17$}}
\put( -.9933,1.510){\makebox(0,0){$18$}}
\put( 1.6967,3.060){\makebox(0,0){$19$}}
\put( 1.0667,3.372){\makebox(0,0){$20$}}
\put( -.2733,1.610){\makebox(0,0){$21$}}
\put( -.1833,2.602){\makebox(0,0){$22$}}
\put( -.3700,3.712){\makebox(0,0){$23$}}
\put(  .8267,3.122){\makebox(0,0){$24$}}
\put( -.7400,3.576){\makebox(0,0){$25$}}
\put( 1.4133,2.780){\makebox(0,0){$26$}}
\put(  .1433,3.126){\makebox(0,0){$27$}}
\put(  .0367,2.542){\makebox(0,0){$28$}}
\put( -.1267,1.752){\makebox(0,0){$29$}}
\put(  .2300,2.028){\makebox(0,0){$30$}}
\put( 1.8100,3.512){\makebox(0,0){$31$}}
\put( -.2033,3.360){\makebox(0,0){$32$}}
\put( -.0800,2.528){\makebox(0,0){$33$}}
\put( -.5967,3.396){\makebox(0,0){$34$}}
\put( -.4300,4.480){\makebox(0,0){$35$}}
\put(  .2200,1.788){\makebox(0,0){$36$}}
\put(  .1367,3.746){\makebox(0,0){$37$}}
\put( -.4500,1.884){\makebox(0,0){$38$}}
\put( -.8000,3.744){\makebox(0,0){$39$}}
\put(  .9000,3.264){\makebox(0,0){$40$}}
\put( 1.7900,3.080){\makebox(0,0){$41$}}
\put( 1.0500,2.446){\makebox(0,0){$42$}}
\put(-1.3500,1.758){\makebox(0,0){$43$}}
\put( -.2467,2.242){\makebox(0,0){$44$}}
\put( -.5100,3.980){\makebox(0,0){$45$}}
\put( -.0967,2.364){\makebox(0,0){$46$}}
\put(  .0200,4.012){\makebox(0,0){$47$}}
\put( -.0167,2.624){\makebox(0,0){$48$}}
\put( -.1367,2.738){\makebox(0,0){$49$}}
\put(-1.4467,3.614){\makebox(0,0){$50$}}
\put( -.0967,4.314){\makebox(0,0){$51$}}
\put(  .7333,2.208){\makebox(0,0){$52$}}
\put(  .2600,1.974){\makebox(0,0){$53$}}
\put(  .3767,1.632){\makebox(0,0){$54$}}
\put( -.4767,2.768){\makebox(0,0){$55$}}
\put( -.8567,1.934){\makebox(0,0){$56$}}
\put( -.2667,2.824){\makebox(0,0){$57$}}
\put( 1.5800,3.044){\makebox(0,0){$58$}}
\put(  .2233,3.076){\makebox(0,0){$59$}}
\put( -.0133,2.128){\makebox(0,0){$60$}}
\put( -.0733,2.724){\makebox(0,0){$61$}}
\put( -.0267,3.920){\makebox(0,0){$62$}}
\put( -.4600,3.562){\makebox(0,0){$63$}}
\put(  .6700,2.506){\makebox(0,0){$64$}}
\put( -.6100,5.440){\makebox(0,0){$65$}}
\put( 1.3000,3.740){\makebox(0,0){$66$}}
\put( -.2533,3.162){\makebox(0,0){$67$}}
\put( -.9267,3.282){\makebox(0,0){$68$}}
\put(  .8133,4.126){\makebox(0,0){$69$}}
\put(  .9933,3.590){\makebox(0,0){$70$}}
\put(-1.2300,3.158){\makebox(0,0){$71$}}
\put( -.3833,3.508){\makebox(0,0){$72$}}
\put( 1.9567,3.906){\makebox(0,0){$73$}}
\put(-1.2400,2.530){\makebox(0,0){$74$}}
\put( -.4167,3.368){\makebox(0,0){$75$}}
\put(-1.2667,1.432){\makebox(0,0){$76$}}
\put(  .9167,3.268){\makebox(0,0){$77$}}
\put(  .3967,2.874){\makebox(0,0){$78$}}
\put(  .2067,5.948){\makebox(0,0){$79$}}
\put(  .5633,2.632){\makebox(0,0){$80$}}
\put( -.0600,4.974){\makebox(0,0){$81$}}
\put(  .4333,2.460){\makebox(0,0){$82$}}
\put( 3.3067,6.254){\makebox(0,0){$83$}}
\put( -.3267,3.364){\makebox(0,0){$84$}}
\put(-1.0567,2.996){\makebox(0,0){$85$}}
\put( -.5833,3.274){\makebox(0,0){$86$}}
\put( 1.1800,4.836){\makebox(0,0){$87$}}
\put(  .7967,3.652){\makebox(0,0){$88$}}
\put(-1.1833,4.016){\makebox(0,0){$89$}}
\put( -.0467,2.820){\makebox(0,0){$90$}}
\put(  .1800,1.680){\makebox(0,0){$91$}}
\put( 1.5567,3.742){\makebox(0,0){$92$}}
\put(  .1700,2.180){\makebox(0,0){$93$}}
\put(  .1867,4.954){\makebox(0,0){$94$}}
\put(  .9367,4.450){\makebox(0,0){$95$}}
\put( -.5133,1.272){\makebox(0,0){$96$}}
\put( -.5533,3.542){\makebox(0,0){$97$}}
\put( 3.4033,5.476){\makebox(0,0){$98$}}
\end{picture}
\end{center}
\caption{A scatter plot of spring (MAM) precipitation in De Bilt,
Netherlands versus the NINO3 index of SST in the eastern Pacific for
1857--1998.  Underlined numbers refer to the 19th
century.  The horizontal thin lines give the 10\%, 33\%, 67\% and 90\%
percentiles.}
\label{fig:scatter}
\end{figure*}

\begin{figure*}
\begin{center}
\setlength{\unitlength}{0.1bp}
\begin{picture}(3600,2160)(0,0)
\put(0,0){\psfig{file=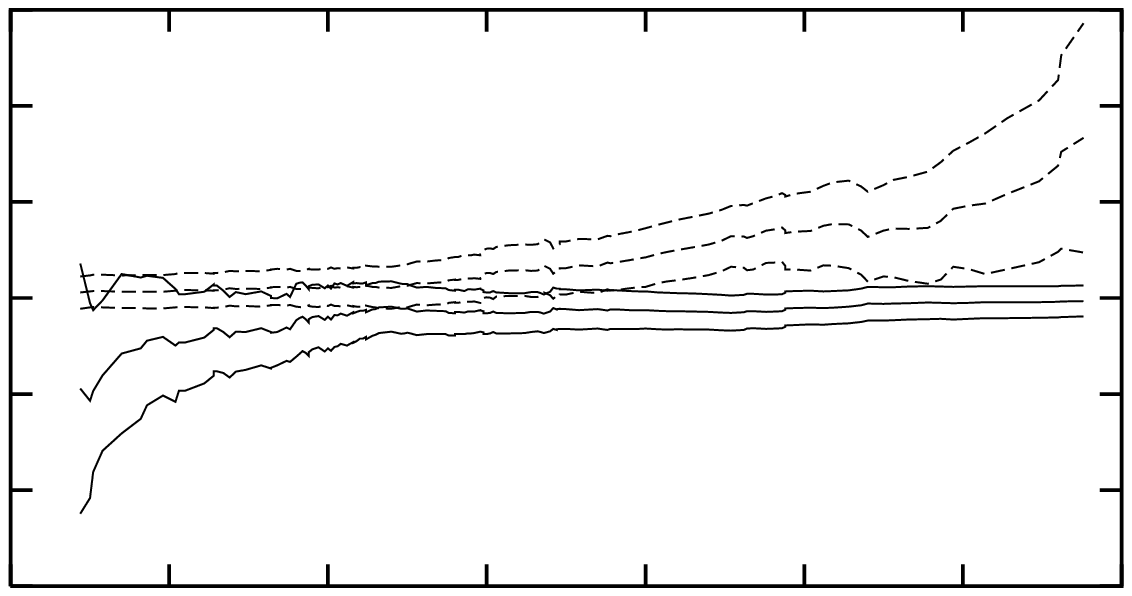}}
\put(3550,50){\makebox(0,0)[tr]{$N_3^\mathrm{cut}$}}
\put(1950,2060){\makebox(0,0){De Bilt}}
\put(0,2010){\makebox(0,0)[bl]{MAM prec.}}
\put(3550,150){\makebox(0,0){$2$}}
\put(3093,150){\makebox(0,0){$1.5$}}
\put(2636,150){\makebox(0,0){$1$}}
\put(2179,150){\makebox(0,0){$0.5$}}
\put(1721,150){\makebox(0,0){$0$}}
\put(1264,150){\makebox(0,0){$-0.5$}}
\put(807,150){\makebox(0,0){$-1$}}
\put(350,150){\makebox(0,0){$-1.5$}}
\put(300,1910){\makebox(0,0)[r]{$300$}}
\put(300,1357){\makebox(0,0)[r]{$200$}}
\put(300,803){\makebox(0,0)[r]{$100$}}
\put(300,250){\makebox(0,0)[r]{$0$}}
\end{picture}
\end{center}
\caption{Mean and $2\sigma$ uncertainties of the precipitation in De Bilt 
for the years with $N_3 > N_3^\mathrm{cut}$ (dashed curves), and 
$N_3<N_3^\mathrm{cut}$ (solid curves).}
\label{fig:cutmean}
\end{figure*}

To quantize the significance of these relationships
we used a Kolmogorov-Smirnov test to compute the
probability that the precipitation distribution with NINO3 index
$N_3<N_3^\mathrm{cut}$ is the same as the one with
$N_3>N_3^\mathrm{cut}$.  The averages and $2\sigma$ bands of these 
distributions of the De Bilt data are shown in Figure~\ref{fig:cutmean}.  
The difference is significant at the 99\% level for $N_3^\mathrm{cut} > 
0.5$: an El Ni\~no tends to be followed by a wet spring.  
On the other hand, the effects of La Ni\~na, including the 
suggestive drought in 1893, are not significant even at the 95\% 
confidence level at De Bilt.  For the four-station average this
difference is also significant at the 95\% level for
$N_3^\mathrm{cut}<-0.5$.

The signal has no connection with the North Atlantic Oscillation
(NAO).  On the one hand the NAO does not influence precipitation in
the Netherlands very much, on the other hand ENSO and NAO are only
correlated in the summer.


\section{Spring rain in Europe}
\label{sec:europe}

\begin{figure*}
\begin{center}
\psfig{file=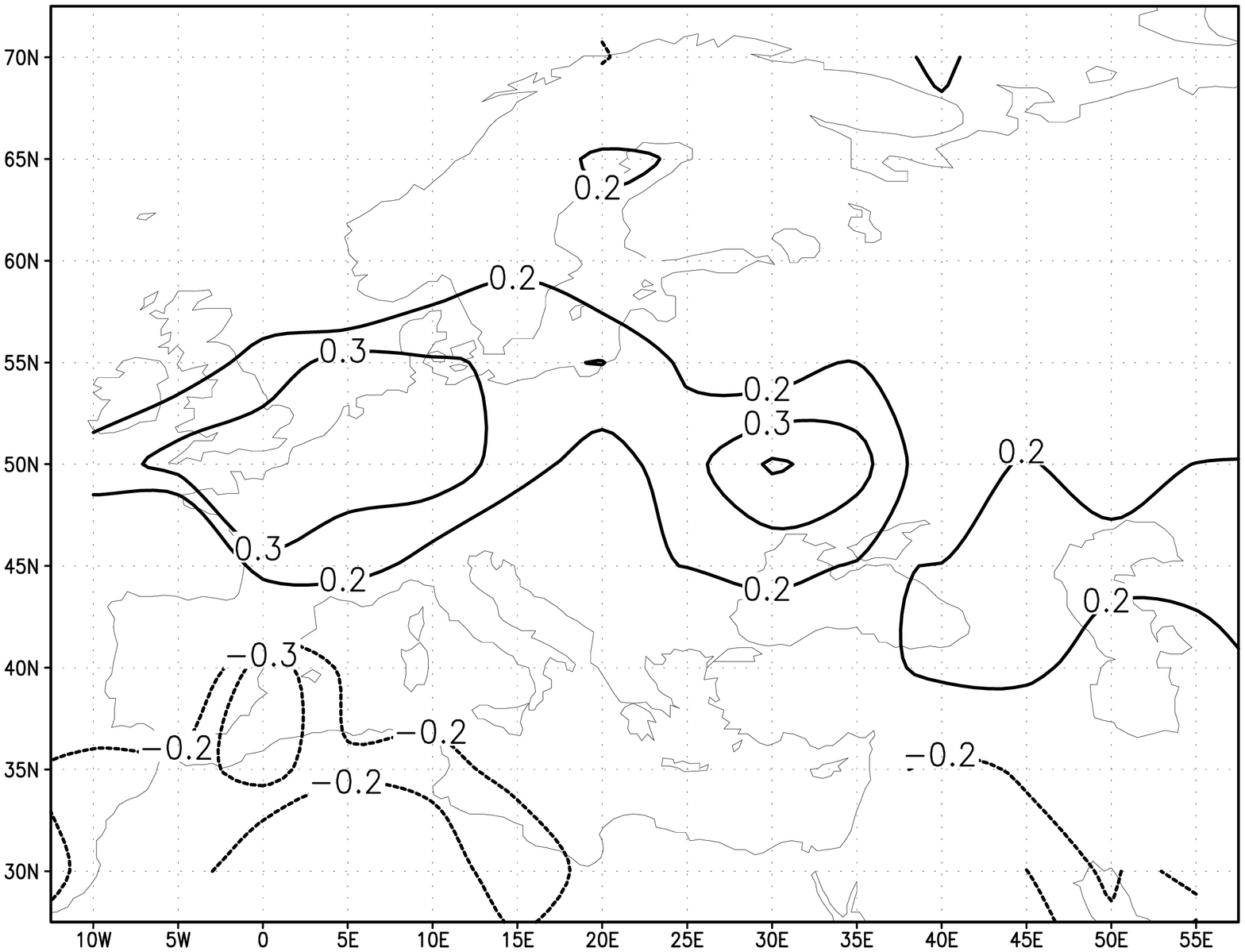,angle=-90,width=12cm}
\end{center}
\caption{The linear correlation coefficients of the DJF NINO3 index and the 
MAM precipitation over Europe (1857--1993).}
\label{fig:europe}
\end{figure*}

We investigated the extent of the teleconnection using the NCDC
gridded precipitation anomalies database \citep{NCDCprcp1995}.  This
contains global data from 1851 to 1993 in 5\dg$\times$5\dg\ bins.  In
Figure~\ref{fig:europe} one sees that the spring precipitation
increases after an El Ni\~no in a zonal belt from England and France
to the Ukraine, with a weaker extension eastwards into Asia.  There
are two maxmima, with correlations coefficients above $0.3$ (without
the 1998 El Ni\~no): southern England, northern France, the Low
Countries and Germany, and another one in the Ukraine centered on Kiev
($r=0.43$).  This last point was also noted by
\citet{KiladisDiaz1989}.  In their Fig.~3h a similar but more
southerly band is indicated over Europe.  This band forms a dipole
with the drier zone over Northern Africa and eastern Spain
($r=-0.35$), also noted in \citet{KiladisDiaz1989}.  In contrast, the
correlation of DJF precipitation with the DJF NINO3 index only reaches
values above $0.2$ in three grid points: $0.22$ in Brussels, $0.23$ in
Moscow and $-0.29$ in Bergen, Norway; none of these reach 99\%
significance.  The Iberian signals in the summer and early fall are
also weaker.

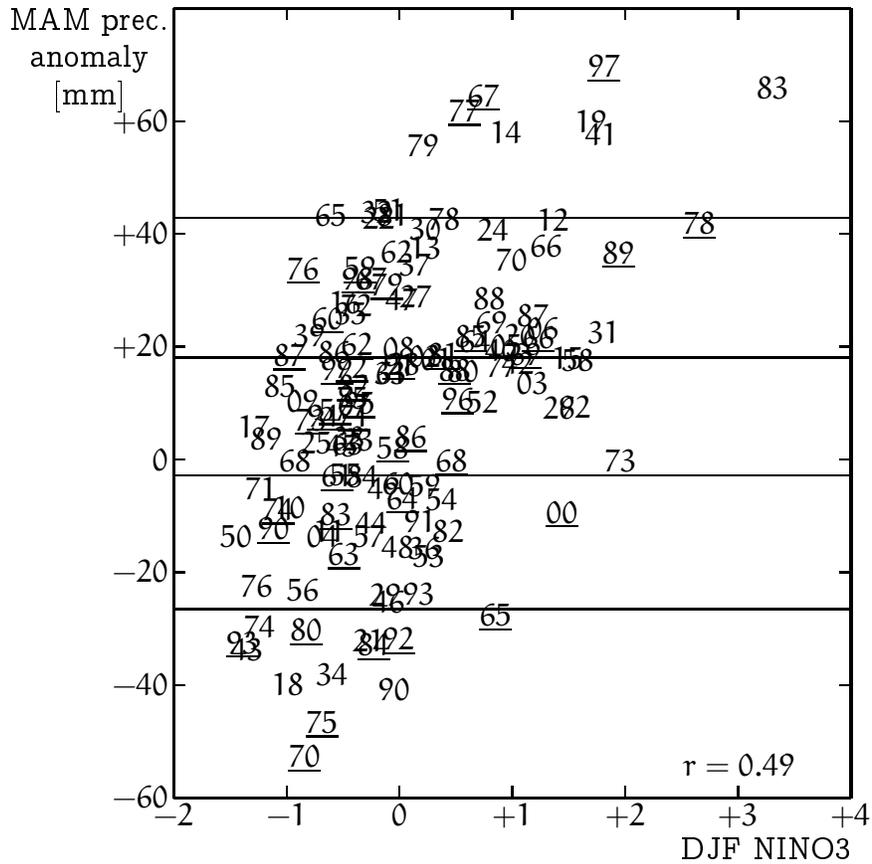
\begin{figure}
\begin{center}
\unitlength 1.5cm
\begin{picture}(6.4,7.4)(-2.4,-3.4)
\thicklines
\put(-2,-3){\line(1,0){6}}
\put(-2,-3){\line(0,1){7}}
\put(-2,+4){\line(1,0){6}}
\put(4,-3){\line(0,1){7}}
\put(3.5,-2.8){\makebox(0,0)[br]{$r=0.49$}}
\thinlines 
\put(-2,-1.330){\line(1,0){6}}
\put(-2,-.1415){\line(1,0){6}}
\put(-2,.90475){\line(1,0){6}}
\put(-2,2.1385){\line(1,0){6}}
\put(-1,-3){\line(0,1){.10}}
\put( 0,-3){\line(0,1){.10}}
\put( 1,-3){\line(0,1){.10}}
\put( 2,-3){\line(0,1){.10}}
\put( 3,-3){\line(0,1){.10}}
\put(-1,4){\line(0,-1){.10}}
\put( 0,4){\line(0,-1){.10}}
\put( 1,4){\line(0,-1){.10}}
\put( 2,4){\line(0,-1){.10}}
\put( 3,4){\line(0,-1){.10}}
\put(-2,-2){\line(1,0){.10}}
\put(-2,-1){\line(1,0){.10}}
\put(-2,0){\line(1,0){.10}}
\put(-2,1){\line(1,0){.10}}
\put(-2,2){\line(1,0){.10}}
\put(-2,3){\line(1,0){.10}}
\put(4,-2){\line(-1,0){.10}}
\put(4,-1){\line(-1,0){.10}}
\put(4,0){\line(-1,0){.10}}
\put(4,1){\line(-1,0){.10}}
\put(4,2){\line(-1,0){.10}}
\put(4,3){\line(-1,0){.10}}
\put(-2,-3.05){\makebox(0,0)[t]{$-2$}}
\put(-1,-3.05){\makebox(0,0)[t]{$-1$}}
\put( 0,-3.05){\makebox(0,0)[t]{$0$}}
\put( 1,-3.05){\makebox(0,0)[t]{$+1$}}
\put( 2,-3.05){\makebox(0,0)[t]{$+2$}}
\put( 3,-3.05){\makebox(0,0)[t]{$+3$}}
\put( 4,-3.05){\makebox(0,0)[t]{$+4$}}
\put( 4,-3.35){\makebox(0,0)[tr]{DJF NINO3}}
\put(-2.05,-3){\makebox(0,0)[r]{$-60$}}
\put(-2.05,-2){\makebox(0,0)[r]{$-40$}}
\put(-2.05,-1){\makebox(0,0)[r]{$-20$}}
\put(-2.05,0){\makebox(0,0)[r]{$0$}}
\put(-2.05,1){\makebox(0,0)[r]{$+20$}}
\put(-2.05,2){\makebox(0,0)[r]{$+40$}}
\put(-2.05,3){\makebox(0,0)[r]{$+60$}}
\put(-2.05,4){\makebox(0,0)[rt]{\shortstack{MAM prec.\\anomaly\\{}[mm]}}}
\put( -0.5733,  0.4180){\makebox(0,0){$\underline{57}$}}
\put( -0.0600,  0.0880){\makebox(0,0){$\underline{58}$}}
\put( -0.3500,  1.6760){\makebox(0,0){$\underline{59}$}}
\put( -0.6400,  1.2315){\makebox(0,0){$\underline{60}$}}
\put( -0.5533, -0.1670){\makebox(0,0){$\underline{61}$}}
\put( -0.3767,  0.9980){\makebox(0,0){$\underline{62}$}}
\put( -0.4933, -0.8615){\makebox(0,0){$\underline{63}$}}
\put(  0.0267, -0.3650){\makebox(0,0){$\underline{64}$}}
\put(  0.8533, -1.4000){\makebox(0,0){$\underline{65}$}}
\put(  1.2300,  1.0645){\makebox(0,0){$\underline{66}$}}
\put(  0.7433,  3.2075){\makebox(0,0){$\underline{67}$}}
\put(  0.4600, -0.0310){\makebox(0,0){$\underline{68}$}}
\put(  1.1133,  0.9135){\makebox(0,0){$\underline{69}$}}
\put( -0.8400, -2.6525){\makebox(0,0){$\underline{70}$}}
\put( -0.4033,  0.3665){\makebox(0,0){$\underline{71}$}}
\put( -0.4200,  0.7990){\makebox(0,0){$\underline{72}$}}
\put( -0.7867,  0.3310){\makebox(0,0){$\underline{73}$}}
\put( -1.0767, -0.4650){\makebox(0,0){$\underline{74}$}}
\put( -0.6867, -2.3515){\makebox(0,0){$\underline{75}$}}
\put( -0.8567,  1.6735){\makebox(0,0){$\underline{76}$}}
\put(  0.5767,  3.0735){\makebox(0,0){$\underline{77}$}}
\put(  2.6567,  2.0725){\makebox(0,0){$\underline{78}$}}
\put( -0.1100,  1.5300){\makebox(0,0){$\underline{79}$}}
\put( -0.8267, -1.5330){\makebox(0,0){$\underline{80}$}}
\put(  0.3867,  0.9080){\makebox(0,0){$\underline{81}$}}
\put( -0.4033,  0.6295){\makebox(0,0){$\underline{82}$}}
\put( -0.5633, -0.5130){\makebox(0,0){$\underline{83}$}}
\put( -0.2300, -1.6670){\makebox(0,0){$\underline{84}$}}
\put(  0.6300,  1.0655){\makebox(0,0){$\underline{85}$}}
\put(  0.1000,  0.1805){\makebox(0,0){$\underline{86}$}}
\put( -0.9733,  0.9015){\makebox(0,0){$\underline{87}$}}
\put(  0.4900,  0.7725){\makebox(0,0){$\underline{88}$}}
\put(  1.9400,  1.8135){\makebox(0,0){$\underline{89}$}}
\put( -1.1133, -0.6355){\makebox(0,0){$\underline{90}$}}
\put( -0.0033,  0.8165){\makebox(0,0){$\underline{91}$}}
\put( -0.0100, -1.6160){\makebox(0,0){$\underline{92}$}}
\put( -1.3933, -1.6430){\makebox(0,0){$\underline{93}$}}
\put( -0.6767,  0.3685){\makebox(0,0){$\underline{94}$}}
\put( -0.3567,  0.4790){\makebox(0,0){$\underline{95}$}}
\put(  0.5167,  0.5150){\makebox(0,0){$\underline{96}$}}
\put(  1.8133,  3.4650){\makebox(0,0){$\underline{97}$}}
\put( -0.3733,  1.5850){\makebox(0,0){$\underline{98}$}}
\put( -0.5567,  0.7740){\makebox(0,0){$\underline{99}$}}
\put(  1.4400, -0.4900){\makebox(0,0){$\underline{00}$}}
\put(  0.3267,  0.8895){\makebox(0,0){$01$}}
\put(  0.2233,  0.9135){\makebox(0,0){$02$}}
\put(  1.1767,  0.6830){\makebox(0,0){$03$}}
\put( -0.6833, -0.6765){\makebox(0,0){$04$}}
\put(  0.9467,  1.0105){\makebox(0,0){$05$}}
\put(  1.2667,  1.1670){\makebox(0,0){$06$}}
\put( -0.4033,  0.4815){\makebox(0,0){$07$}}
\put( -0.0067,  0.9870){\makebox(0,0){$08$}}
\put( -0.8600,  0.5205){\makebox(0,0){$09$}}
\put( -0.9733, -0.4245){\makebox(0,0){$10$}}
\put( -0.6200, -0.6310){\makebox(0,0){$11$}}
\put(  1.3600,  2.1225){\makebox(0,0){$12$}}
\put(  0.2300,  1.8780){\makebox(0,0){$13$}}
\put(  0.9433,  2.8980){\makebox(0,0){$14$}}
\put(  1.4900,  0.9015){\makebox(0,0){$15$}}
\put( -0.4833,  1.3920){\makebox(0,0){$16$}}
\put( -1.2933,  0.2945){\makebox(0,0){$17$}}
\put( -0.9933, -1.9930){\makebox(0,0){$18$}}
\put(  1.6967,  2.9980){\makebox(0,0){$19$}}
\put(  1.0667,  1.1045){\makebox(0,0){$20$}}
\put( -0.2733, -1.6055){\makebox(0,0){$21$}}
\put( -0.1833,  2.1400){\makebox(0,0){$22$}}
\put( -0.3700,  0.1810){\makebox(0,0){$23$}}
\put(  0.8267,  2.0380){\makebox(0,0){$24$}}
\put( -0.7400,  0.1490){\makebox(0,0){$25$}}
\put(  1.4133,  0.4600){\makebox(0,0){$26$}}
\put(  0.1433,  1.4430){\makebox(0,0){$27$}}
\put(  0.0367,  0.8470){\makebox(0,0){$28$}}
\put( -0.1267, -1.2020){\makebox(0,0){$29$}}
\put(  0.2300,  2.0445){\makebox(0,0){$30$}}
\put(  1.8100,  1.1250){\makebox(0,0){$31$}}
\put( -0.2033,  2.1915){\makebox(0,0){$32$}}
\put( -0.0800,  0.7655){\makebox(0,0){$33$}}
\put( -0.5967, -1.9020){\makebox(0,0){$34$}}
\put( -0.4300,  1.2930){\makebox(0,0){$35$}}
\put(  0.2200, -0.7850){\makebox(0,0){$36$}}
\put(  0.1367,  1.7295){\makebox(0,0){$37$}}
\put( -0.4500,  0.1925){\makebox(0,0){$38$}}
\put( -0.8000,  1.1055){\makebox(0,0){$39$}}
\put(  0.9000,  0.9740){\makebox(0,0){$40$}}
\put(  1.7900,  2.8810){\makebox(0,0){$41$}}
\put(  1.0500,  0.8585){\makebox(0,0){$42$}}
\put( -1.3500, -1.6835){\makebox(0,0){$43$}}
\put( -0.2467, -0.5660){\makebox(0,0){$44$}}
\put( -0.5100,  0.1270){\makebox(0,0){$45$}}
\put( -0.0967, -1.2600){\makebox(0,0){$46$}}
\put(  0.0200,  1.4230){\makebox(0,0){$47$}}
\put( -0.0167, -0.7730){\makebox(0,0){$48$}}
\put( -0.1367, -0.2585){\makebox(0,0){$49$}}
\put( -1.4467, -0.6820){\makebox(0,0){$50$}}
\put( -0.0967,  2.2140){\makebox(0,0){$51$}}
\put(  0.7333,  0.5090){\makebox(0,0){$52$}}
\put(  0.2600, -0.8540){\makebox(0,0){$53$}}
\put(  0.3767, -0.3530){\makebox(0,0){$54$}}
\put( -0.4767, -0.1335){\makebox(0,0){$55$}}
\put( -0.8567, -1.1580){\makebox(0,0){$56$}}
\put( -0.2667, -0.6795){\makebox(0,0){$57$}}
\put(  1.5800,  0.8850){\makebox(0,0){$58$}}
\put(  0.2233, -0.2260){\makebox(0,0){$59$}}
\put( -0.0133, -0.2120){\makebox(0,0){$60$}}
\put( -0.0733,  0.7685){\makebox(0,0){$61$}}
\put( -0.0267,  1.8390){\makebox(0,0){$62$}}
\put( -0.4600,  0.1440){\makebox(0,0){$63$}}
\put(  0.6700,  1.0505){\makebox(0,0){$64$}}
\put( -0.6100,  2.1690){\makebox(0,0){$65$}}
\put(  1.3000,  1.8980){\makebox(0,0){$66$}}
\put( -0.2533,  1.6025){\makebox(0,0){$67$}}
\put( -0.9267, -0.0065){\makebox(0,0){$68$}}
\put(  0.8133,  1.2140){\makebox(0,0){$69$}}
\put(  0.9933,  1.7720){\makebox(0,0){$70$}}
\put( -1.2300, -0.2555){\makebox(0,0){$71$}}
\put( -0.3833,  1.3645){\makebox(0,0){$72$}}
\put(  1.9567, -0.0125){\makebox(0,0){$73$}}
\put( -1.2400, -1.4800){\makebox(0,0){$74$}}
\put( -0.4167,  0.5505){\makebox(0,0){$75$}}
\put( -1.2667, -1.1195){\makebox(0,0){$76$}}
\put(  0.9167,  0.8285){\makebox(0,0){$77$}}
\put(  0.3967,  2.1370){\makebox(0,0){$78$}}
\put(  0.2067,  2.7920){\makebox(0,0){$79$}}
\put(  0.5633,  0.7860){\makebox(0,0){$80$}}
\put( -0.0600,  2.1740){\makebox(0,0){$81$}}
\put(  0.4333, -0.6315){\makebox(0,0){$82$}}
\put(  3.3067,  3.3015){\makebox(0,0){$83$}}
\put( -0.3267, -0.1495){\makebox(0,0){$84$}}
\put( -1.0567,  0.6495){\makebox(0,0){$85$}}
\put( -0.5833,  0.9505){\makebox(0,0){$86$}}
\put(  1.1800,  1.2765){\makebox(0,0){$87$}}
\put(  0.7967,  1.4275){\makebox(0,0){$88$}}
\put( -1.1833,  0.1900){\makebox(0,0){$89$}}
\put( -0.0467, -2.0365){\makebox(0,0){$90$}}
\put(  0.1800, -0.5450){\makebox(0,0){$91$}}
\put(  1.5567,  0.4645){\makebox(0,0){$92$}}
\put(  0.1700, -1.1915){\makebox(0,0){$93$}}
\end{picture}
\end{center}
\caption{Scatter plot of the MAM precipitation anomalies in Europe 
around 50\dg N against the DJF NINO3 index for 1857--1993.}
\label{fig:index_scatter}
\end{figure}

\begin{figure}
\begin{center}
\setlength{\unitlength}{0.1bp}
\begin{picture}(3600,2160)(0,0)
\put(0,0){\psfig{file=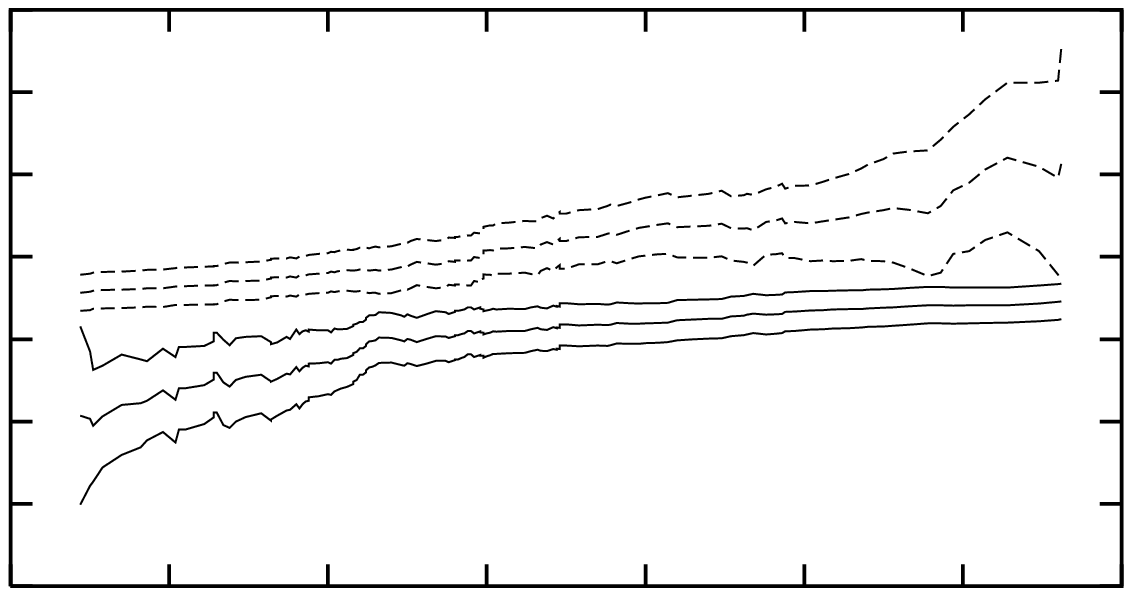}}
\put(1950,2060){\makebox(0,0){Europe 50\dg N index}}
\put(3550,50){\makebox(0,0)[tr]{$N_3^\mathrm{cut}$}}
\put(3550,150){\makebox(0,0){$2$}}
\put(3093,150){\makebox(0,0){$1.5$}}
\put(2636,150){\makebox(0,0){$1$}}
\put(2179,150){\makebox(0,0){$0.5$}}
\put(1721,150){\makebox(0,0){$0$}}
\put(1264,150){\makebox(0,0){$-0.5$}}
\put(807,150){\makebox(0,0){$-1$}}
\put(350,150){\makebox(0,0){$-1.5$}}
\put(0,2010){\makebox(0,0)[bl]{MAM prec.\ anom.}}
\put(300,1910){\makebox(0,0)[r]{$80$}}
\put(300,1436){\makebox(0,0)[r]{$40$}}
\put(300,961){\makebox(0,0)[r]{$0$}}
\put(300,487){\makebox(0,0)[r]{$-40$}}
\end{picture}
\end{center}
\caption{Mean and $2\sigma$ uncertainties of the precipitation anomaly 
around 50\dg N from 5\dg W to 35\dg E.  Dashed curves: the
years with $N_3 > N_3^\mathrm{cut}$, solid curves: below
$N_3<N_3^\mathrm{cut}$.}
\label{fig:50Ncutmean}
\end{figure}

The relation with ENSO is seen more clearly when we construct an index of
average MAM rainfall anomalies over the band with positive correlations 
consisting of the nine 5\dg$\times$5\dg\ grid boxes centered on 50\dg N 
from 5\dg W to 35\dg E{}.  
This index has a correlation coefficient with DJF NINO3 of $r=0.49$, 
and one can see from figure~\ref{fig:50Ncutmean} that the effect now 
looks significant for all values of $N_3$.  A K-S test confirms that 
the distributions are different for the entire range of cut-off
values.  The correlation with the NAO is again low, although non-zero
($r=-0.16$).

\section{Possible mechanisms}
\label{sec:mechanisms}

\begin{figure}
\begin{center}
\setlength{\unitlength}{0.1bp}
\begin{picture}(3600,2160)(0,0)
\put(0,0){\psfig{file=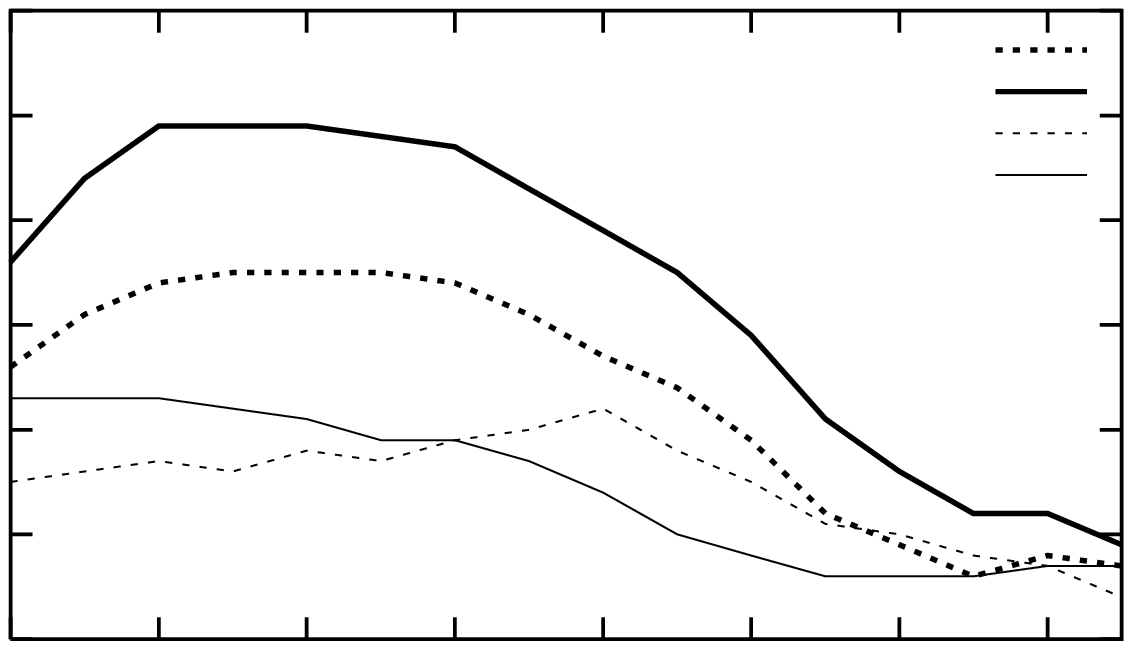}}
\put(3137,1587){\makebox(0,0)[r]{Europe 50\dg N index DJF}}
\put(3137,1707){\makebox(0,0)[r]{De Bilt DJF}}
\put(3137,1827){\makebox(0,0)[r]{Europe 50\dg N index MAM}}
\put(3137,1947){\makebox(0,0)[r]{De Bilt MAM}}
\put(3550,50){\makebox(0,0)[tr]{lag [months]}}
\put(100,2060){\makebox(0,0)[tr]{$r$}}
\put(3337,150){\makebox(0,0){$14$}}
\put(2910,150){\makebox(0,0){$12$}}
\put(2483,150){\makebox(0,0){$10$}}
\put(2057,150){\makebox(0,0){$8$}}
\put(1630,150){\makebox(0,0){$6$}}
\put(1203,150){\makebox(0,0){$4$}}
\put(777,150){\makebox(0,0){$2$}}
\put(350,150){\makebox(0,0){$0$}}
\put(300,2060){\makebox(0,0)[r]{$0.6$}}
\put(300,1758){\makebox(0,0)[r]{$0.5$}}
\put(300,1457){\makebox(0,0)[r]{$0.4$}}
\put(300,1155){\makebox(0,0)[r]{$0.3$}}
\put(300,853){\makebox(0,0)[r]{$0.2$}}
\put(300,552){\makebox(0,0)[r]{$0.1$}}
\put(300,250){\makebox(0,0)[r]{$0.0$}}
\end{picture}
\end{center}
\caption{Lag correlation coefficients of the precipitation in spring 
and winter in De Bilt and in Europe around 50\dg N with the NINO3
index.}
\label{fig:lagcor}
\end{figure}

\begin{figure*}
\figwidth=.31\textwidth
\psfig{file=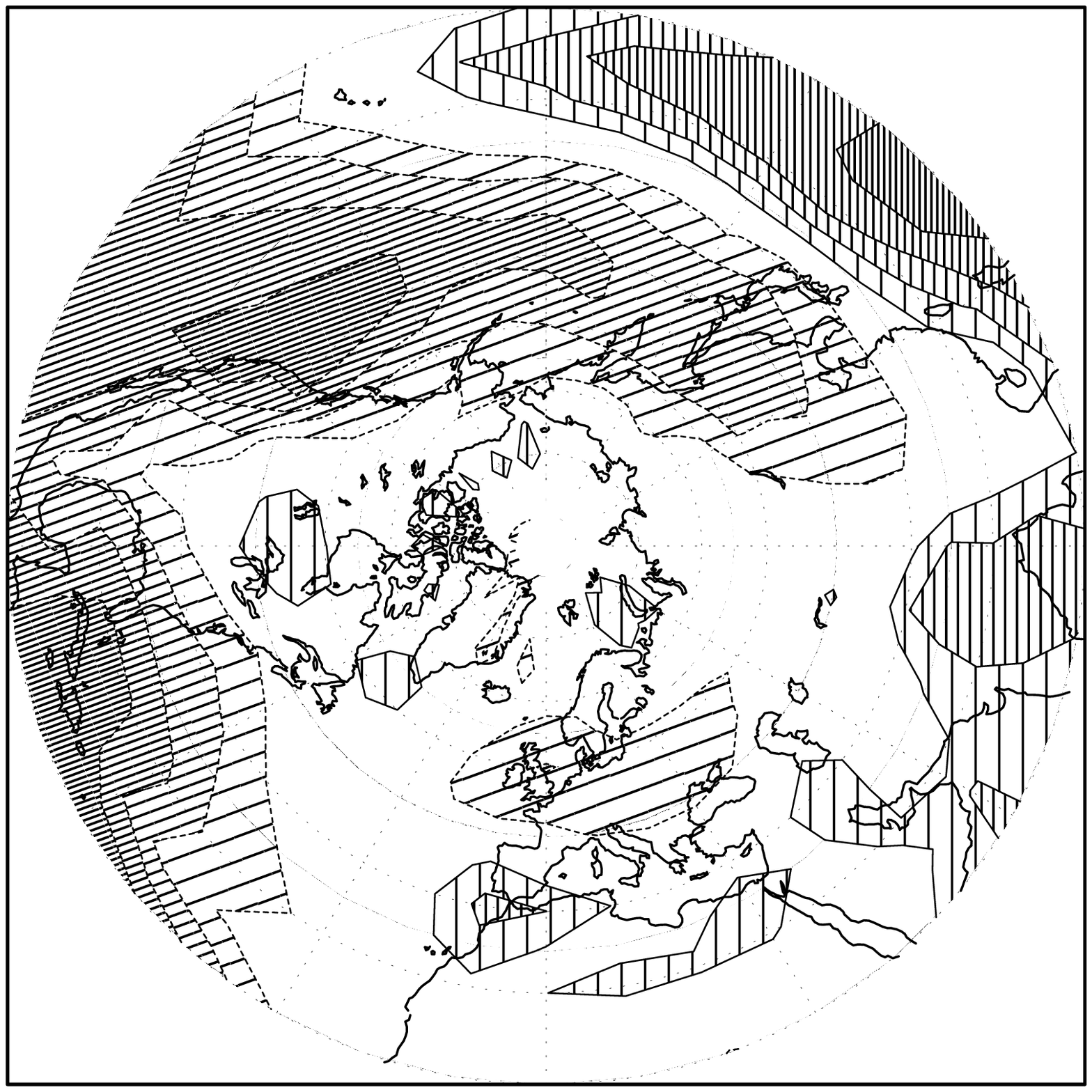,angle=-90,width=\figwidth}
\psfig{file=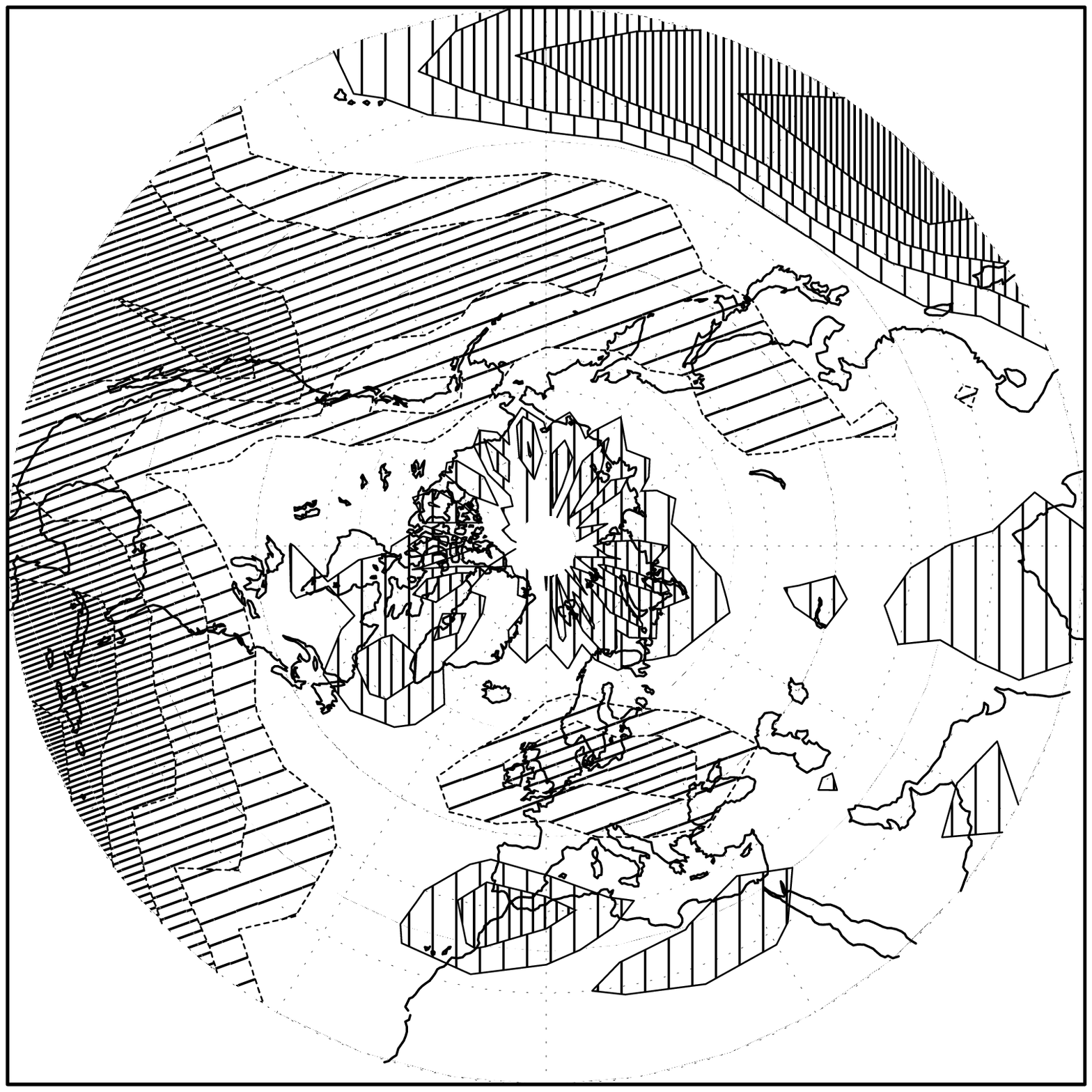,angle=-90,width=\figwidth}
\psfig{file=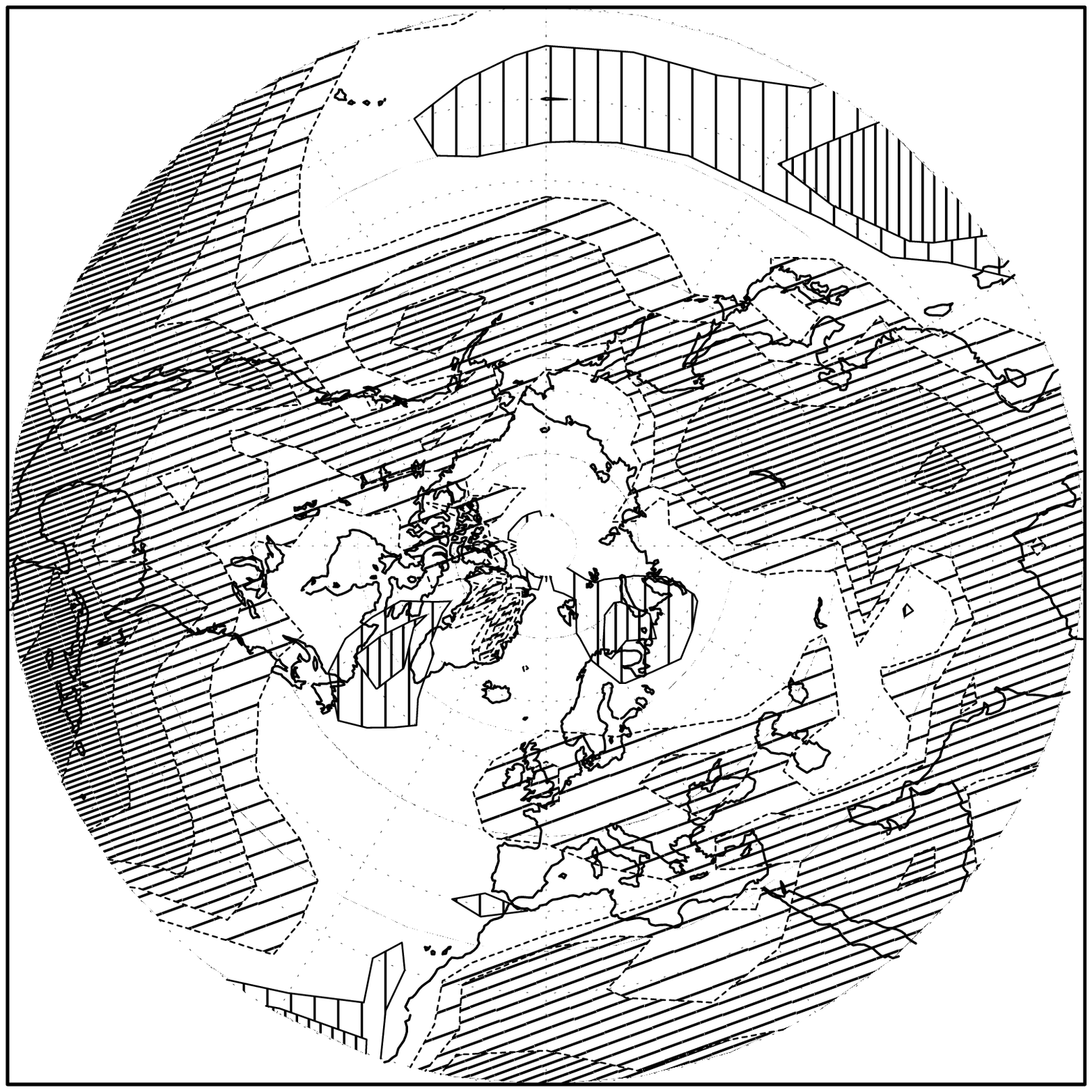,angle=-90,width=\figwidth}
\psfig{file=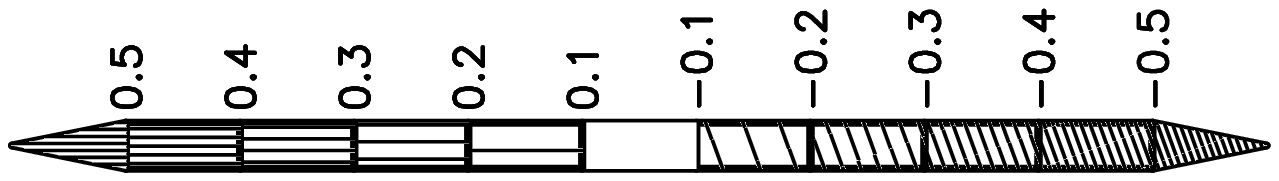,angle=-90,width=.125\figwidth}
\caption{Correlation maps of MAM northern hemisphere SLP (1873-1995) and
MAM NINO3 (a), DJF NINO3 (b), and an index of SE Asia SST (c).}
\label{fig:SLP}
\end{figure*}

Possible mechanisms of this teleconnection have to explain the time
structure of the correlation.  The lag correlations of the MAM
precipitation with the NINO3 index is shown in
figure~\ref{fig:lagcor}.  For reference we also show the DJF
correlations.  Although there is room for an atmospheric mechanism
with a time scale shorter than a month, the main signal seems to be
delayed by 3--6 months.  This agrees with the observation that the
correlations of the MAM NINO3 index with historical sea level pressure
data \citep[1873--1995, ][]{Jones1987,BasnellParker1997} are not very
high, though significant.  In Fig.~\ref{fig:SLP}a one sees that in a band
from the British Isles to the Ukraine sea-level pressure tends to be
lower during El-Ni\~no events.  The correlation coefficient 
$r$ just reaches $-0.20$ ($P=97\%$) over the North Sea and
$-0.16$ in the Ukraine ($P=92\%$).  It is on average a somewhat
higher in Northern Africa, $r=0.23$ at the Straits of Gibraltar
($P=99\%$).  This is the correct structure to explain more rain in the
dipole of figure~\ref{fig:europe}.  The lag-3 signal
(Fig.~\ref{fig:SLP}b) has the same features, but is stronger:
$r=-0.26$ over the North Sea, $0.27$ at Gibraltar.

\begin{figure*}
\figwidth=.86\textwidth
\begin{center}
{$r_{\mathrm{SST},P(\mathrm{50^\circ N})}$}\\
\vspace{-3mm}
\psfig{file=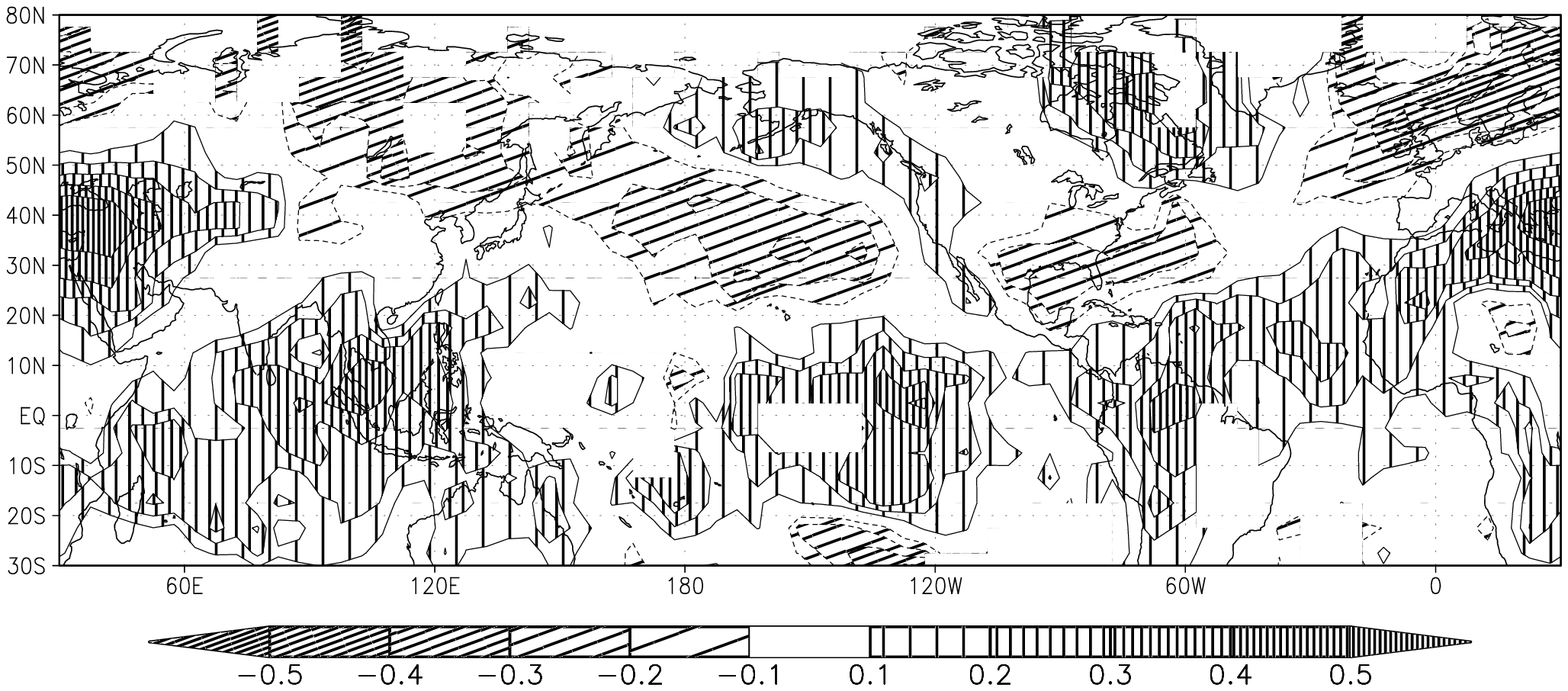,angle=-90,width=\figwidth}\\
{$r_{\mathrm{NINO3},\mathrm{SST}}^{\mathrm{lag}3}$}\\
\vspace{-3mm}
\psfig{file=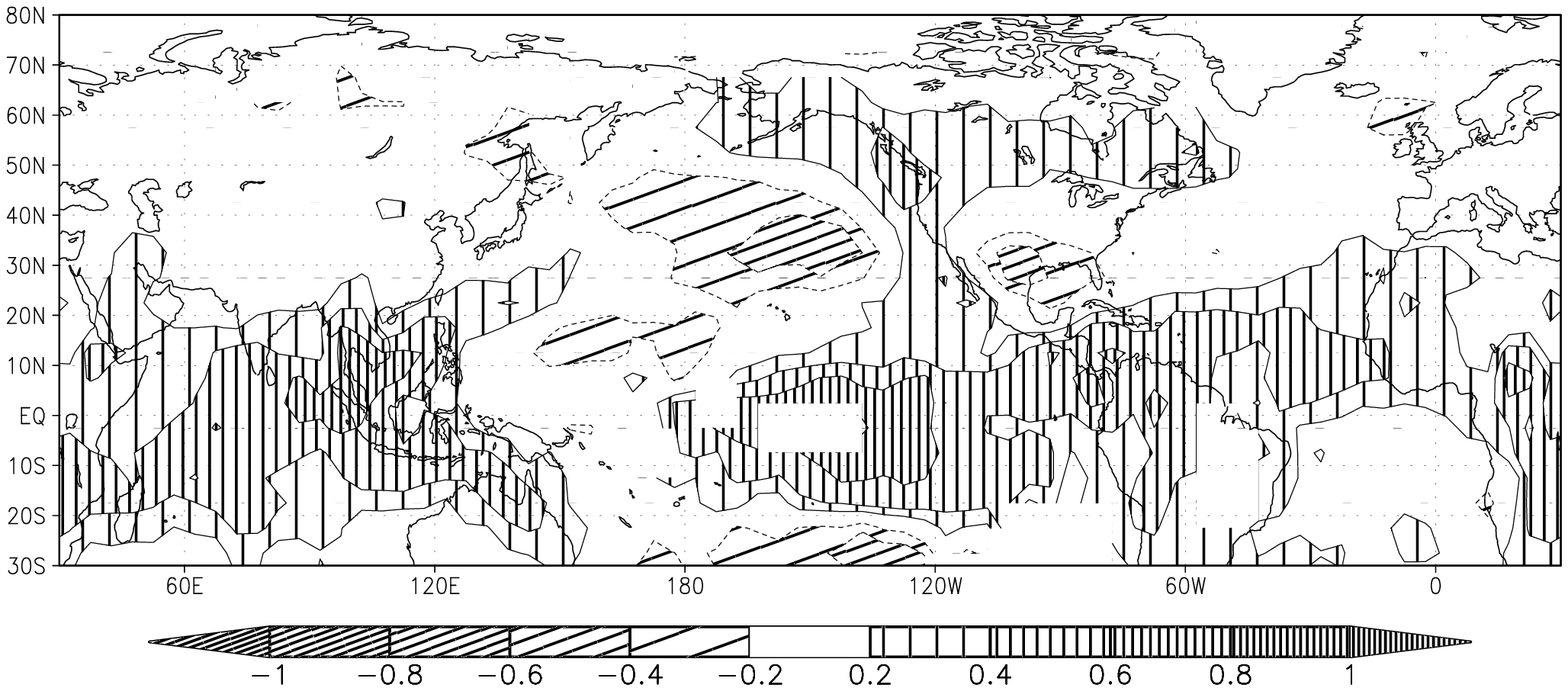,angle=-90,width=\figwidth}\\
{$r_{\mathrm{NINO3},\mathrm{SST}}^{\mathrm{lag}3}
\times r_{\mathrm{SST},P(\mathrm{50^\circ N})}$}\\
\vspace{-3mm}
\psfig{file=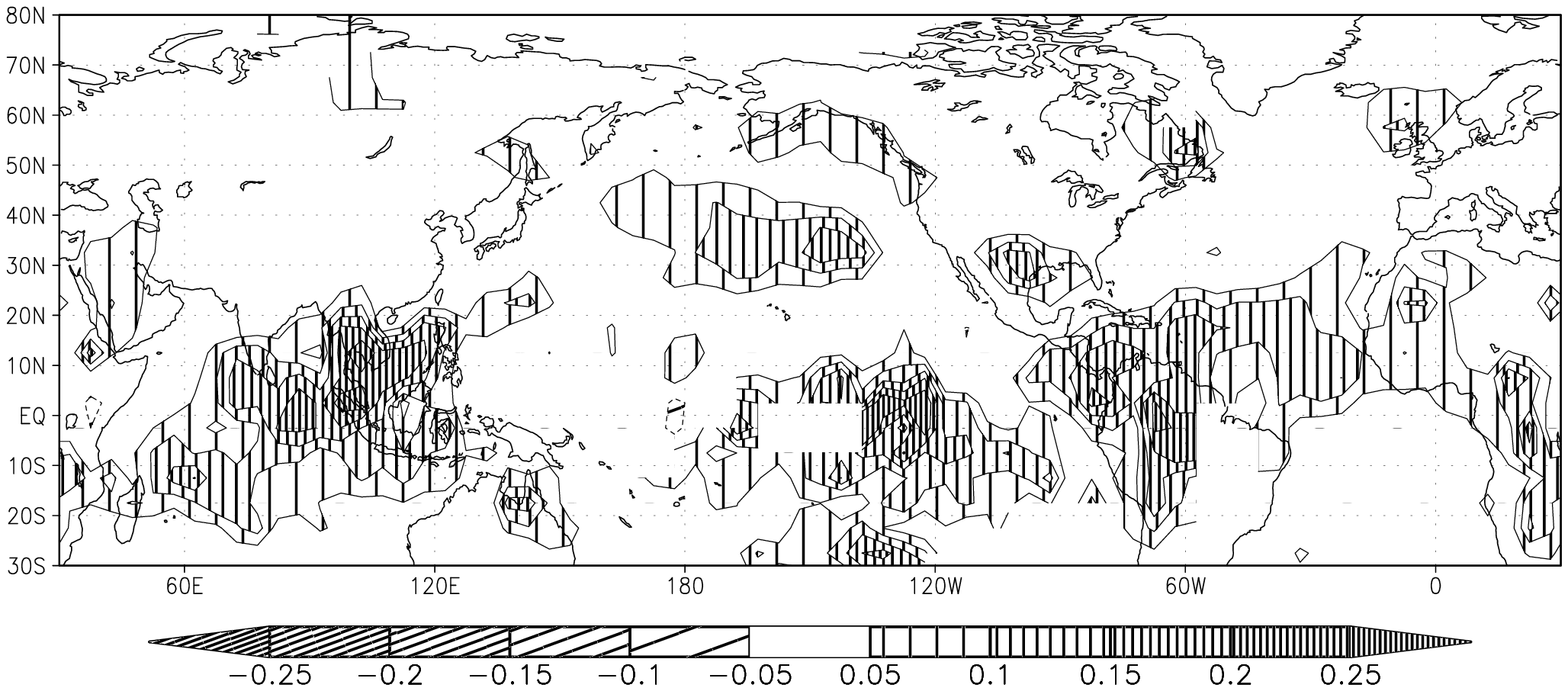,angle=-90,width=\figwidth}
\end{center}
\caption{The correlation of spring precipitation around 50\dg N in 
Europe with the Jones and Parker temperature dataset (top), the lag-3  
correlation of this temperature and the NINO3 index (middle) and the 
product of the these two correlations (bottom).  Note the different 
scales.}
\label{fig:rr}
\end{figure*}

The 3--6 month delay points at the possibility that in addition to
this direct teleconnection there is an
intermediate variable, probably SST in a third region, that is
influenced by ENSO and in turn causes more rain around 50\dg N in
Europe in spring.  To investigate this we use the historical temperature
anomalies database of Jones and Parker
\citep{Parker1995,Jones1994,Parker1994}, which includes SST as well as
land 2~m temperatures.  In Fig.~\ref{fig:rr} (top panel) we show the 
correlation of these temperatures with the European 50\dg N spring
precipitation index.  Locally, high precipitation is associated with
colder water in the north-east Atlantic.  One also recognizes the NAO
SST signature in the West Atlantic, in spite of the low correlation
with the atmospheric NAO index.  However, both these patterns are only
very weakly associated with ENSO, as one can see from the middle panel
in which the correlations of this MAM temperature field and the DJF
NINO3 index are shown.

The overlap between these two plots is shown in the bottom panel,
in which the product of the top two panels is plotted,
$r_{\mathrm{NINO3},\mathrm{SST}}^{\mathrm{lag}3} \times
r_{\mathrm{SST},P(\mathrm{50^\circ N})}$.  If only one area would act
as intermediate variable the local value would be equal to
$r_{\mathrm{NINO3},P(\mathrm{50^\circ N})}^{\mathrm{lag}3} = 0.49$.  
Even if more intermediate variables contribute, areas in
which both correlations are high will stand out in this plot, but a 
quantitative interpretation cannot be given.  One sees that none of the
regions reach values as high as $0.49$, but there are 
three areas of possible interest: the central equatorial Pacific,
south-east Asia and parts of the Indian Ocean, and the North Pacific
dipole.

\begin{table}
\begin{tabular}{rr|cccccc}
    &       & DJF   & MAM     &         &           &         &         \\
    &       & NINO3 & NINO3.4 & SE Asia & N Pacific & De Bilt & 50\dg N \\ 
\hline
DJF & NINO3 & 1.00  & 0.80    & 0.67    & 0.49      & 0.35    & 0.49 \\
MAM & NINO3.4 &     & 1.00    & 0.67    & 0.50      & 0.22    & 0.28 \\
    & SE Asia &     &         & 1.00    & 0.47      & 0.36    & 0.35 \\
    & N Pacific &   &         &         & 1.00      & 0.26    & 0.30 \\
    & De Bilt &     &         &         &           & 1.00    & 0.60 \\
    & 50\dg N &     &         &         &           &         & 1.00 \\
\end{tabular}
\caption{Correlation coefficients between the DJF Kaplan/NCEP NINO3
index; the MAM NINO3.4, SE Asia and North Pacific temperature indices 
extracted from \citet{Parker1995}; and MAM De Bilt and European 50\dg
N precipitation.}
\label{tab:correlations}
\end{table}

\begin{figure}
\begin{center}
\setlength{\unitlength}{0.1bp}
\begin{picture}(3600,2160)(0,0)
\put(0,0){\psfig{file=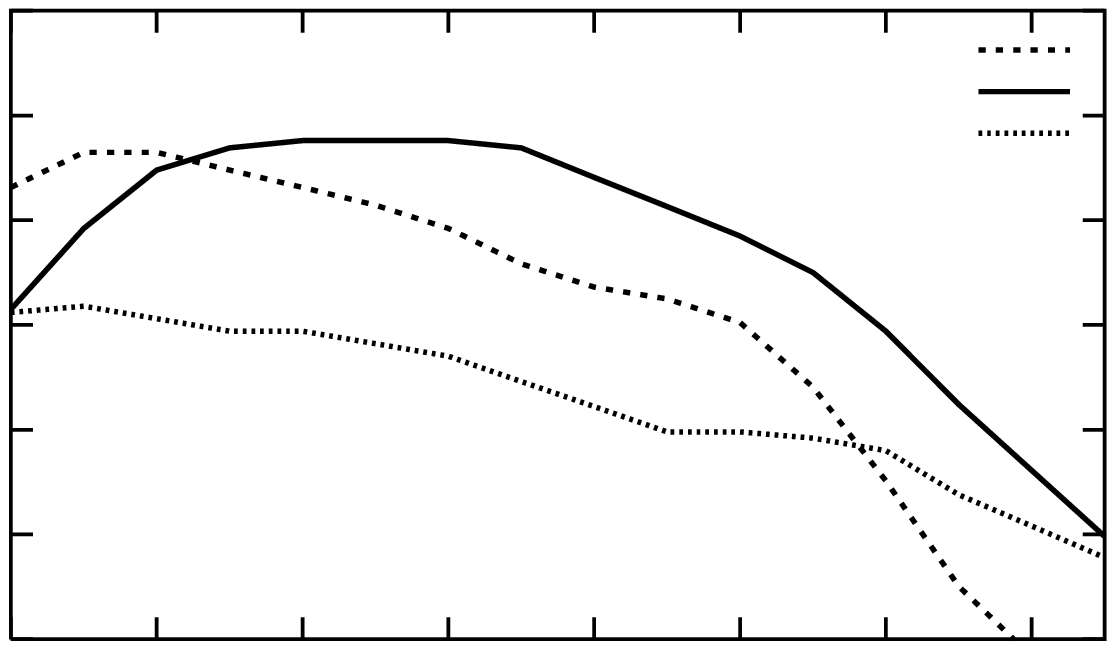}}
\put(3550,50){\makebox(0,0)[tr]{lag [months]}}
\put(100,2060){\makebox(0,0)[tr]{$r$}}
\put(3137,1707){\makebox(0,0)[r]{$0.30\times\mbox{}$N Pacific}}
\put(3137,1827){\makebox(0,0)[r]{$0.35\times\mbox{}$SE Asia}}
\put(3137,1947){\makebox(0,0)[r]{$0.28\times\mbox{}$NINO3.4}}
\put(3340,150){\makebox(0,0){$14$}}
\put(2920,150){\makebox(0,0){$12$}}
\put(2500,150){\makebox(0,0){$10$}}
\put(2080,150){\makebox(0,0){$8$}}
\put(1660,150){\makebox(0,0){$6$}}
\put(1240,150){\makebox(0,0){$4$}}
\put(820,150){\makebox(0,0){$2$}}
\put(400,150){\makebox(0,0){$0$}}
\put(350,2060){\makebox(0,0)[r]{$0.3$}}
\put(350,1457){\makebox(0,0)[r]{$0.2$}}
\put(350,853){\makebox(0,0)[r]{$0.1$}}
\put(350,250){\makebox(0,0)[r]{$0.0$}}
\end{picture}
\end{center}
\caption{Lag correlation coefficients of the MAM temperature anomalies in the 
three possible source regions with the with the NINO3 index, weighted with 
the correlation coefficients with the Europe 50\dg N precipitation
index in spring.}
\label{fig:lagcor2}
\end{figure}

We defined three temperature anomaly indices corresponding to these
regions: for the Central Pacific we use the NINO3.4 region, for
south-east Asia the box 60\dg E -- 120\dg E, 10\dg S -- 20\dg N and
for the North Pacific dipole the region 160\dg E -- 120\dg W, 30\dg N
-- 60\dg N with a top-left/bottom-right dipole structure.  The
correlation coefficients between all indices we accumulated are given
in Table~\ref{tab:correlations}. In figure~\ref{fig:lagcor2} we plot
the lag-correlation coefficients of these indices with the NINO3
index, multiplied by their correlation with the European 50\dg N
spring precipitation index.  One sees that the shape of the lag
structure of the south-east Asian region corresponds closely to the
observed signal.  Also the signal in pressure (Fig.~\ref{fig:SLP}c) is
very similar to the lagged NINO3 signal over Europe
(Fig.~\ref{fig:SLP}b), but the North-African opposite-sign anomaly has
disappeared.  However, the correlation of the south-east Asia index
with rainfall in Europe is only $0.35$, which is lower than the lagged
correlation with NINO3 instead of the higher value expected for an
intermediate variable.  This may partly be due to other mechanisms,
such as the direct link from the central Pacific, but also the nature
of the measurements plays a role.  Temperature differences in the
south-east Asia area are small: the MAM variance is only
$(0.28\:\mathrm{K})^2$, much less than the $(0.88\:\mathrm{K})^2$ of
the DJF NINO3 index.  This increases the effect of noise in the form
of measurement errors, irrelevant land points and small-scale weather.
On the basis of the time delays of the signal we conclude that
south-east Asia is most likely the main source of the influence of
ENSO on the weather of Europe, with a smaller direct link from the
central Pacific.

The importance of the south-east Asia sea surface temperature for the
northern hemisphere circulation is supported by data analyses and
modelling studies, see e.g.\ the review by \citet{TrenberthTOGA}.  In
spring the most active teleconnection is the North Pacific pattern.
This is the top half of Figs~\ref{fig:SLP}a,b.  The lower halves show
an extension across the North Pole into Europe.  This extension
substantiates arguments using simplified Rossby wave propagation.
Unfortunately virtually all modelling studies have been conducted for
northern winter conditions, when the observations contain a much
weaker teleconnection.  Still, indications of the pole-crossing
response can be seen in DJF AGCM results \citep[see,
e.g.,][]{Ferranti1994}.  One could speculate that the spring
transition to the Asian and Chinese monsoon systems makes the
circulation more susceptible to perturbations.  Further modelling work
is clearly needed in order to elucidate the mechanism behind the
teleconnection to Europe.


\section{Conclusions}

Using more than a century of data a clear influence of ENSO on the
weather in Europe has been established: spring precipitation in a belt
around 50\dg N from Southern England to the Ukraine tends to increase
after an El Ni\~no and decrease after a La Ni\~na.  The strength of
the correlation is $r=0.49$ for an index of precipitation in this
belt, $r=0.40$ for the average of four Dutch stations, and $r=0.35$
for the single station De Bilt.  Other European teleconnections are
weaker than this.  The 3-6 month lag and correlation maps suggest that
south-east Asian surface temperatures may act as an intermediate
variables for most of the signal.

\paragraph{Acknowledgements}\hspace*{-5pt} We would like to thank
Theo Opsteegh and Robert Mureau for many useful discussions.



\begin{thebibliography}{19}
\expandafter\ifx\csname natexlab\endcsname\relax\def\natexlab#1{#1}\fi

\bibitem[Allan \emph{et~al.}(1991)Allan, Nicholls, Jones, and
  Butterworth]{Allan1991}
Allan, R.J., Nicholls, N., Jones, P.D., and Butterworth, I.J. 1991. `A further
  extension of the {Tahiti}-{Darwin} {SOI}, early {SOI} results and {Darwin}
  pressure',
\newblock {\em J.\ Climate,} {\bf 4,} 743--749.
\newblock The series can be found at
  http://\discretionary{}{}{}www.cru.uea.ac.uk/\discretionary{}{}{}cru/data/so%
i.htm.

\bibitem[Baker \emph{et~al.}(1995)Baker, Eischeid, Karl, and
  Diaz]{NCDCprcp1995}
Baker, C.~B., Eischeid, J.~K., Karl, T.~R., and Diaz, H.~F. 1995.
\newblock `The quality control of long-term climatological data using objective
  data analysis',
\newblock in {\em Preprints of AMS Ninth Conference on Applied Climatology}
  Dallas, TX.
\newblock The data are available from
  http://www.ncdc.noaa.gov/\discretionary{}{}{}onlinedata/\discretionary{}{}{}%
climatedata/\discretionary{}{}{}grid.prcp.seasanom.html.

\bibitem[Basnett and Parker(1997)]{BasnellParker1997}
Basnett, T.~A. and Parker, D.~E. 1997.
\newblock `Development of the global mean sea level pressure data set
  {GMSLP2}',
\newblock Climatic Research Technical Note~79 Hadley Centre Meteorological
  Office, Bracknell, U.K.
\newblock 16pp plus Appendices. Data are available from
  {}http://\discretionary{}{}{}www.cru.uea.ac.uk/\discretionary{}{}{}cru/data/%
\discretionary{}{}{}pressure.htm.

\bibitem[Berlage(1966)]{Berlage1966}
Berlage, H.~P. 1966.
\newblock {\em The Southern Oscillation and World Weather}
\newblock Number~88 in Mededelingen en verhandelingen. KNMI 152 pp.

\bibitem[Ferranti \emph{et~al.}(1994)Ferranti, Molteni, and
  Palmer]{Ferranti1994}
Ferranti, L., Molteni, F., and Palmer, N. 1994. `Impact of localized tropical
  and extratropical {SST} anomalies in ensembles of seasonal {GCM}
  integrations',
\newblock {\em Q.\ J.\ Meteorol.\ Soc.,} {\bf 120,} 1613--1645.

\bibitem[Fraedrich(1994)]{Fraedrich1994}
Fraedrich, K. 1994. `An {ENSO} impacty on {Europe?}',
\newblock {\em Tellus,} {\bf 46A,} 541--552.

\bibitem[Gray(1984)]{Gray1984a}
Gray, W.~M. 1984. `Atlantic seasonal hurricane frequency: {Part I}: {El
  Ni\~{n}o} and 30 mb quasi--biennial oscillation influences',
\newblock {\em Mon.\ Wea.\ Rev.,} {\bf 112,} 1649--1668.

\bibitem[Halpert and Ropelewski(1992)]{HalpertRopelewski1992}
Halpert, M.~S. and Ropelewski, C.~F. 1992. `Surface temperature patterns
  associated with the {Southern Oscillation}',
\newblock {\em J.\ Climate,} {\bf 5,} 577--593.

\bibitem[Jones(1987)]{Jones1987}
Jones, P.~D. 1987. `The early twentieth century arctic high --- fact or
  fiction?',
\newblock {\em Climate Dynamics,} {\bf 1,} 63--75.

\bibitem[Jones(1994)]{Jones1994}
Jones, P.~D. 1994. `Hemispheric surface air temperature variations: a
  reanalysis and an update to 1993',
\newblock {\em J.\ Climate,} {\bf 7,} 1794--1802.

\bibitem[Kaplan \emph{et~al.}(1998)Kaplan, Cane, Kushnir, Clement, Blumenthal,
  and Rajagopalan]{Kaplan1998}
Kaplan, A., Cane, M., Kushnir, Y., Clement, A.~C., Blumenthal, and Rajagopalan,
  B. 1998. `Analyses of global sea surface temperature 1856--1991',
\newblock {\em J.\ Geophys.\ Res.,} {\bf 103,} 18567--18589.
\newblock Data are available from
  {http://\discretionary{}{}{}ingrid.ldgo.columbia.edu.}

\bibitem[Kiladis and Diaz(1989)]{KiladisDiaz1989}
Kiladis, G.~N. and Diaz, H.~F. 1989. `Global climatic anomalies associated with
  extremes in the southern oscillation',
\newblock {\em J.\ Climate,} {\bf 2,} 1069--1090.

\bibitem[K\"onnen \emph{et~al.}(1998)K\"onnen, Jones, Kaltofen, and
  Allan]{Konnen1998}
K\"onnen, G.~P., Jones, P.~D., Kaltofen, M.~H., and Allan, R.~J. 1998.
  `Pre-1866 extensions of the {Southern Oscillations} index using early
  {Indonesian} and {Tahitian} meteorological readings',
\newblock {\em J.\ Climate,} {\bf 11,} 2325--2339.

\bibitem[Parker \emph{et~al.}(1995)Parker, Folland, and Jackson]{Parker1995}
Parker, D.~E., Folland, C.~K., and Jackson, M. 1995. `Marine surface
  temperature: observed variations and data requirements',
\newblock {\em Climatic Change,} {\bf 31,} 559--600.

\bibitem[Parker \emph{et~al.}(1994)Parker, Jones, Bevan, and
  Folland]{Parker1994}
Parker, D.~E., Jones, P.~D., Bevan, A., and Folland, C.~K. 1994. `Interdecadal
  changes of surface temperature since the late 19th century',
\newblock {\em J.\ Geophys.\ Res.,} {\bf 99,} 14373--14399.
\newblock Data are available from
  http://\discretionary{}{}{}www.cru.uea.ac.uk/\discretionary{}{}{}cru/data/te%
mperat.htm.

\bibitem[Reynolds and Smith(1994)]{ReynoldsAnalyses}
Reynolds, R.~W. and Smith, T.~M. 1994. `Improved global sea surface analyses
  using optimum interpolation',
\newblock {\em J.\ Clim.,} {\bf 7,} 929--948.
\newblock \textsc{Nino} indices are available from the {Climate Prediction
  Center} at
  http://nic.fb4.noaa.gov/\discretionary{}{}{}data/cddb/\discretionary{}{}{}al%
tindex.html.

\bibitem[Trenberth \emph{et~al.}(1998)Trenberth, Branstator, Karoly, Kumar,
  Lau, and Ropelewski]{TrenberthTOGA}
Trenberth, K.~E., Branstator, G.~W., Karoly, D., Kumar, A., Lau, N.-C., and
  Ropelewski, C. 1998. `Progress during {TOGA} in understanding and modeling
  global teleconnections associated with tropical sea surface temperatures',
\newblock {\em J.\ Geophys. Res.,} {\bf 103,} 14291--14324.

\bibitem[van Loon and Madden(1981)]{VanLoonMadden1981}
van Loon, H. and Madden, R.~A. 1981. `The {Southern Oscillation}, {Part I}.
  {G}lobal associations with pressure and temperature in northern winter',
\newblock {\em Mon.\ Wea.\ Rev.,} {\bf 109,} 1150--1162.

\bibitem[Wilby(1993)]{Wilby1993}
Wilby, R. 1993. `Evidence of {ENSO} in the synoptic climate of the {British}
  {Isles} since 1880',
\newblock {\em Weather,} {\bf 48,} 234--239.

\end{thebibliography}

\end{document}